\documentclass[a4paper,11pt]{article}
\usepackage{jcappub} % for details on the use of the package, please
\usepackage[T1]{fontenc} % if needed

\usepackage{color, colortbl}
\definecolor{LightCyan}{rgb}{0.88,1,1}
\definecolor{Gray}{gray}{0.9}

\usepackage[utf8]{inputenc}
\title{\boldmath Theoretical foundations of the reduced relativistic gas in the cosmological perturbed context}

\author[a,1]{G. Pordeus-da-Silva}
\author[b]{R. C. Batista,\note{Corresponding author.}}
\author[b]{and L. G. Medeiros,}
\affiliation[a]{Departamento de F\'isica Te\'orica e Experimental - Universidade
	Federal do Rio Grande do Norte, Natal - RN, Brasil.}
	
\affiliation[b]{ Escola de Ci\^{e}ncia e Tecnologia - Universidade Federal do Rio Grande do Norte, Natal - RN, Brasil.}

\emailAdd{givalpordeus@ufrn.edu.br}
\emailAdd{rbatista@ect.ufrn.br}
\emailAdd{leogmedeiros@ect.ufrn.br}

\abstract{The Reduced Relativistic Gas (RRG) is a simplified version of the ideal relativistic gas, which assumes that all particles have the same momentum magnitude. Although this is a very idealized situation, the resulting model preserves the phenomenology of Maxwell-Boltzmann distribution and, in some situations, can be described as a perfect fluid, without introducing large errors in both cosmological background and first-order perturbations. The perfect fluid description of RRG model was already used to study the warmness of dark matter, massive neutrinos and interaction of baryons and photons before recombination, showing very good agreement with previous works based on the full Einstein-Boltzmann system of equations. In order to understand these results and construct a more general and formal framework for RRG, we develop  a theoretical description of first-order cosmological perturbations of RRG, based on a distribution function which encodes the simplifying assumption that all particles have the same momentum magnitude. The full set of Einstein-Boltzmann equations for RRG distribution are derived and quantities beyond the perfect fluid approximation are studied. Using RRG to describe warm dark matter, we show that, for particles with $m \sim \text{keV}$, the perfect fluid approximation is valid on scales $k < 10\, \text{h}/\text{Mpc}$, for most of the universe evolution. We also determine initial conditions for RRG in the early universe and study the evolution of potential in a toy model of universe composed only by RRG. 
}


\arxivnumber{1904.09904}

\begin{document}

\maketitle

\flushbottom

\section{Introduction} \label{intro} 

The ideal relativistic gas is a good description of a real gas when interactions between the constituent particles can be neglected. In the classical regime, when average occupation numbers are much smaller than unity, this model is described by the Maxwell-Boltzmann distribution function \footnote{Frequently, the Maxwell-Boltzmann distribution function is called J\"{o}uttner-Synge distribution function  \cite{mi2011introduction} due to F. J\"{u}ttner's work in 1911 \cite{Juttner1911} and J.L. Synge's in 1957 \cite{synge1957}.}. Despite the simplicity of the model, the resulting equations in cosmological context can not be solved analytically, both in background and first-order perturbations. With a further simplifying condition, that all particles of the relativistic gas have the same momentum magnitude, the relativistic gas and the cosmological equations become simpler, at the cost of an error smaller than $2.5\%$ in the density evolution with respect to Maxwell-Boltzmann description \cite{RRG2005BERREDO-PEIXOTO}. This model is called Reduced Relativistic Gas (RRG), and was first introduced in 1966 by Sakharov for Cosmic Microwave Background applications \cite{RRGsakharov1966}. 

The RRG was recently revived in Ref.~\cite{RRG2005BERREDO-PEIXOTO}, where background analytical solutions for scenarios with radiation and cosmological constant were studied. It was also used to provide a simpler fluid description of Warm Dark Matter (WDM), which, based on linear power spectrum data, produced stringent constraints on the thermal velocities of the particles without the necessity of solving the full set Einstein-Boltzmann equations \cite{RRG2009Fabris}. Moreover, analytical solutions for flat and non-flat background in the presence of other two or three components (relativistic and non-relativistic matter and cosmological constant) were found in Ref.~\cite{RRG2012Leo}. Other uses of RRG include description of barion-photon interaction before recombination \cite{RRG2014Fabris}, massive neutrinos \cite{RRG2018MNRAS}, studies of relativistic matter in anisotropic universes \cite{RRG2018dosReis} and a model independent description of warm dark matter \cite{RRG2018Hipolito}, where it was found that the transfer function for RRG model reproduces the full Einstein-Boltzmann calculation for thermal relics \cite{Bode_2001} with $1\%$ precision.

Given these interesting applications of RRG model in cosmology, it is valuable to study the model in a more general theoretical framework. By studying the RRG beyond the perfect fluid description we can have a better understanding of its approximations and more effects such as dissipation and collisional terms can be introduced. In this paper, we implement the main idea of RRG, i.e., that all particles have same momentum magnitude, in distribution function, both for background and first-order perturbations, and derive the corresponding Einstein-Boltzmann equations. Moreover, we also identify the conditions which reproduce the fluid description of RRG.

The paper is organized as follows. In Sec.~\ref{background}, we present the concepts of RRG, review its background evolution and estimate the mass of WDM particles as a function of RRG warmness parameter. In Sec.~\ref{perturbation}, we derive the first-oder Einstein-Boltzmann system of equations for RRG. In Sec. \ref{application}, we study the warm approximation of RRG, which can be used to study WDM in model independent manner, find the initial conditions for RRG perturbations in the early universe and study a toy model of a universe composed solely by RRG. We make our final comments in Sec. \ref{final_comments}.

\section{Background: concepts and basic equations}\label{background}

The RRG is based on the premise which all particles have the same magnitude for the momentum. Thus, one can reasonably assume its distribution function is proportional to a Dirac delta function in the momentums i.e.
\begin{equation}
f_{\mathcal{R}}\left( p,p_{\mathcal{R}}\right) \equiv Cp\delta \left( p-p_{\mathcal{R}}%
\right) ,  \label{func_distrib_rrg}
\end{equation}%
where $p_{\mathcal{R}}$ sets the magnitude for the momentums and $C$ is a dimensionless constant given by 
\begin{equation}
n_{\mathcal{R}}=\int \frac{d^{3}p}{\left( 2\pi \right) ^{3}}f_{\mathcal{R}} \Rightarrow C=\frac{2\pi ^{2}n_{\mathcal{R}}}{p_{\mathcal{R}}^{3}}\text{.}  \label{caso 22}
\end{equation}%
where $n_{\mathcal{R}}$ is the numerical particle density of RRG. Besides, the energy density $\rho _{\mathcal{R}}$\ and pressure $P_{\mathcal{R}}$ for RRG are given by 
\begin{equation}
\rho _{\mathcal{R}}=\int \frac{d^{3}p}{\left( 2\pi \right) ^{3}}E\left(
p\right) f_{\mathcal{R}} =n_{\mathcal{R}}E\left( p_{\mathcal{R}}\right)
\end{equation}%
and 
\begin{equation}
P_{\mathcal{R}}=\int \frac{d^{3}p}{\left( 2\pi \right) ^{3}}\frac{1}{3}\frac{%
p^{2}}{E\left( p\right) }f_{\mathcal{R}} =\frac{n_{%
\mathcal{R}}}{3}\frac{p_{\mathcal{R}}^{2}}{E\left( p_{\mathcal{R}}\right) }.
\end{equation}%
Combining these equations and using the dispersion relation $E^{2}-p_{\mathcal{R}}%
^{2}=m^{2}$, it is possible to obtain the equation of state of RRG
\begin{equation}
P_{\mathcal{R}}=\frac{\rho _{\mathcal{R}}}{3}\left[ 1-\left( \frac{\rho _{d%
\mathcal{R}}}{\rho _{\mathcal{R}}}\right) ^{2}\right] \text{,}
\label{EoS RRG}
\end{equation}%
where $\rho _{d\mathcal{R}}\equiv n_{\mathcal{R}}m$ is the rest energy density. In the non-relativistic (NR) and ultra-relativistic (UR) limits, we have $\rho _{\mathcal{R}}\simeq \rho _{d\mathcal{R}}\Rightarrow $\ $P_{\mathcal{R}}\simeq 0$\ and $\rho _{\mathcal{R}}\gg \rho _{d\mathcal{R}}$ that $P_{\mathcal{R}}\simeq \rho _{\mathcal{R}}/3$, respectively. It is also possible to isolate $\rho _{\mathcal{R}}$ in the last equation by obtaining%
\begin{equation}
\rho _{\mathcal{R}}=\rho _{d\mathcal{R}}\left[ \frac{3}{2}\left( \frac{P_{%
\mathcal{R}}}{\rho _{d\mathcal{R}}}\right) +\sqrt{\frac{9}{4}\left( \frac{P_{%
\mathcal{R}}}{\rho _{d\mathcal{R}}}\right) ^{2}+1}\right] \text{.}
\label{Rho_RRG}
\end{equation}%
The first to deduce the equation of state of RRG from an assumption about its distribution function were S. C. dos Reis and I. L. Shapiro \cite{RRG2018dosReis}.

As we know, the standard description for a classical relativistic ideal gas of massive particles is given by Maxwell-Boltzmann statistics (MB), whose pressure $P_{\text{MB}}$ and energy density $\rho_{\text{MB}}$ are expressed as \cite{Juttner1911,mi2011introduction} 
\begin{equation}
P_{\text{MB}}=nT\text{ \ \ and \ \ }\rho_{\text{MB}}=3nT+nm\frac{K_{1}\left(
m/T\right) }{K_{2}\left( m/T\right) }\text{,}  \label{P e Rho MB}
\end{equation}
where $n \equiv N/V$ is the numerical particle density, $T$ is the temperature and $K_{1}$ and $K_{2}$ are modified Bessel functions of the second kind. Combining these two equations, we can write 
\begin{equation}
\rho_{\text{MB}}=\rho_{d\text{MB}}\left[ 3\left( \frac{P_{\text{MB}}}{\rho_{d\text{MB}}}%
\right) +\frac{K_{1}\left( \rho_{d\text{MB}}/P_{\text{MB}}\right) }{K_{2}\left( \rho_{d%
\text{MB}}/P_{\text{MB}}\right) }\right] \text{.}  \label{Rho_k1/k2}
\end{equation}

From the thermal velocity definition $v_{\text{th}}\equiv\sqrt{(\left\langle E\right\rangle ^{2}-m^{2})/\left\langle E\right\rangle ^{2}}$ \cite{mi2011introduction}, we obtain for Maxwell-Boltzmann gas the following expression: 
\begin{equation}
v_{\text{thMB}}=\sqrt{1-\left[ \frac{K_{1}\left( \rho_{d\text{MB}}/P_{\text{MB}}\right) 
}{K_{2}\left( \rho_{d\text{MB}}/P_{\text{MB}}\right) }+3\frac{P_{\text{MB}}}{\rho_{d\text{MB}}}%
\right] ^{-2}}\text{.}  \label{Vth_boltzmann}
\end{equation}
On the other hand, due to the peculiar RRG distribution function, its thermal velocity is given by:
\begin{equation}
v_{\text{th}\mathcal{R}}=\sqrt{1-\left( \frac{\rho_{d\mathcal{R}}}{\rho_{%
\mathcal{R}}}\right) ^{2}}=\sqrt{1-4\left( 3\frac{P_{\mathcal{R}}}{\rho_{d%
\mathcal{R}}}+\sqrt{9\left( \frac{P_{\mathcal{R}}}{\rho _{d\mathcal{R}}}%
\right) ^{2}+4}\right) ^{-2}}.  
\label{Vth_RRG}
\end{equation}

\begin{figure}[tbh]
\centering
\includegraphics[scale=1.0]{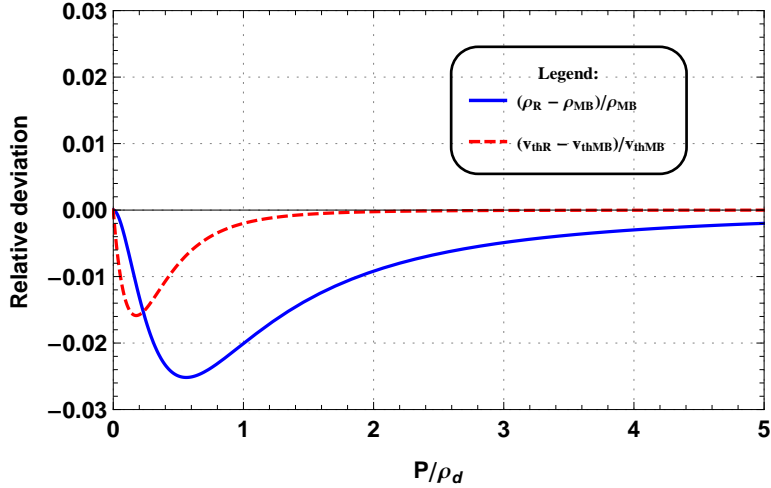}
\caption{Relative deviation of energy density (continuous blue curve) and thermal velocity (dashed red curve) obtained from Maxwell-Boltzmann statistics and RRG model.}
\label{Plot_delta_rho_I}
\end{figure}

Considering that both gases have the same mass and numerical particle density, so that $\rho _{d\mathcal{R}}=\rho _{d\text{MB}}=\rho _{d}$, we can compare them from the functions: 
\begin{equation}
F_{\rho }\equiv \frac{\rho _{\mathcal{R}}-\rho _{\text{MB}}}{\rho _{\text{MB}%
}}\text{ \ \ and \ \ }F_{v_{\text{th}}}\equiv \frac{v_{\text{th}\mathcal{R}%
}-v_{\text{thMB}}}{v_{\text{thMB}}}\text{.}  \label{delta_rho_I}
\end{equation}%
In Fig.~\ref{Plot_delta_rho_I} we show the behavior of these two functions in relation to $P/\rho _{d}$. From this figure, we found that the maximum relative deviation is about $2.5\%$ for the energy density (as was already obtained in Ref.~\cite{RRG2005BERREDO-PEIXOTO}) and about $1.6\%$ for the thermal velocity. We also verified that the relative difference between Maxwell-Boltzmann statistic and RRG model reaches its maximum in a relativistic region where the fraction $P/\rho _{d}$ is close to $1/2$. In the NR ($P/\rho_{d}\simeq 0$) and UR ($P/\rho _{d}\gg 1$) regions, this difference becomes completely negligible.

\subsection{RRG in the cosmological context}\label{RRG no contexto cosmologico}

Now let us consider the evolution of RRG density in Friedmann-Lemaître-Robertson-Walker metric, where $V\propto a^3$. Replacing Eq.~(\ref{EoS RRG}) in conservation law $dU=-PdV$, we obtain 
\begin{equation}
\frac{1}{3}\frac{dV}{V}= -\frac{U_{\mathcal{R}} dU_{\mathcal{R}}} { U_{%
\mathcal{R}}^{2} - U_{d\mathcal{R}}^2}\, ,  \label{energy_cons}
\end{equation}
where $U_{\mathcal{R}}=$ $V\rho_{\mathcal{R}}$ is the total energy and $U_{d\mathcal{R}} = V\rho_{d\mathcal{R}} = \text{const.} $ represents the rest energy density. Integrating Eq.~(\ref{energy_cons}), the evolution of $\rho_{\mathcal{R}}$ with scale factor can be written in the form \cite{RRG2005BERREDO-PEIXOTO,RRG2018Hipolito} 
\begin{equation}
\rho_{\mathcal{R}}\left( a\right) =\rho_{d\mathcal{R},\text{ref}}\left(\frac{a_{\text{ref}}}{a}\right) ^{3}\sqrt{1+b^{2}\left( \frac{a_{\text{ref}}}{a}\right) ^{2}}  \label{eq Rho RRG fuc de a e b1}
\end{equation}
where $b = \sqrt{ U^2_{\mathcal{R},\text{ref}}/ U^2_{d\mathcal{R}} -1}$ is a dimensionless parameter which quantifies the relativistic degree of RRG and is defined at some chosen $a_{\text{ref}}$.

Moreover, the function $\omega\equiv P/\rho$ for RRG is given by 
\begin{equation}
\omega_{\mathcal{R}}\left( a\right) =\frac{1}{3}v_{\text{th}\mathcal{R}}^{2}=%
\frac{1}{3}\left[ \frac{b^{2}}{\left( a/a_{\text{ref}}\right) ^{2}+b^{2}}%
\right] ,  
\label{W RRG}
\end{equation}
where for $a=a_{\text{ref}}$,%
\begin{equation}
b=\frac{v_{\text{th}\mathcal{R},\text{ref}}}{\sqrt{1-v_{\text{th}\mathcal{R},%
\text{ref}}^{2}}}\text{.}
\end{equation}
Observe that $\omega_{\mathcal{R}}\simeq v_{\text{th}\mathcal{R}}\simeq0$ for $a\gg ba_{\text{ref}}$ (NR limit) and for $a\ll ba_{\text{ref}}$ (UR limit) we have $v_{\text{th}\mathcal{R}}^{2}\simeq1$ and $\omega_{\mathcal{R}}\simeq1/3$. Therefore, $b$ is a relative parameter, since this measure is established by comparing the value of the scale factor $a$ with the product of $ba_{\text{ref}}$. Thus, the value of the parameter $b$ that quantify the relativistic degree of the fluid depends on the specific choice made for $a_{\text{ref}}$.

\begin{figure}[th]
\centering
\includegraphics[scale=0.75]{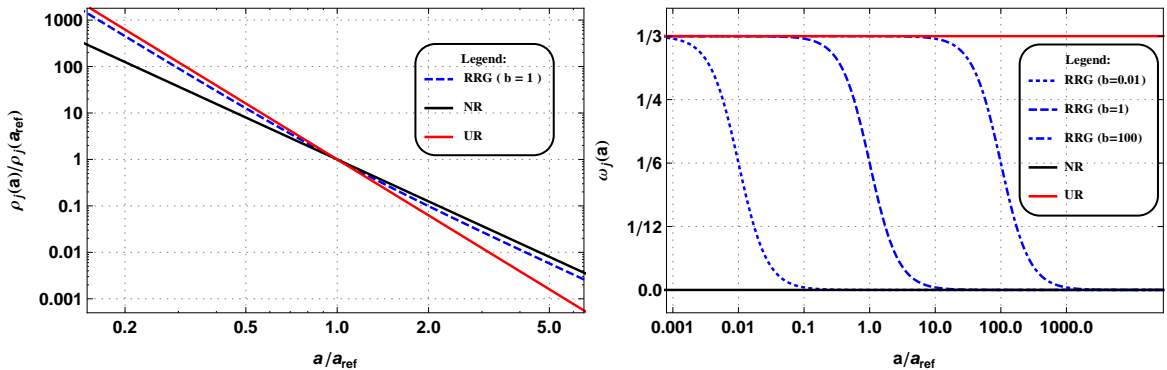}
\caption{Evolution of the energy density $\protect\rho (a)/\protect\rho (a_{\text{ref}})$ (left) and the function $\protect\omega (a)$ (right) with the scale factor for RRG, NR and UR components.}
\label{Fig1a1b}
\end{figure}

For a better interpretation of these relations, see Fig.~\ref{Fig1a1b}. In the figure to the left, we show the dependence between the energy density and the scale factor for NR (continuous black curve), UR (continuous red curve) and RRG (dashed blue curve) components, assuming $b=1$. Note that for $a\ll ba_{\text{ref}}$ and $a\gg ba_{\text{ref}}$ RRG exhibits a typically behavior of UR and NR components, respectively. The $\rho _{\mathcal{R}}\left( a\right) $ interpolates between a typical dependence of a NR component to a UR one at $a\sim ba_{\text{ref}}$. The same conclusion is obtained analyzing the dependence on the scale factor of the function $\omega _{\mathcal{R}}$ in Fig.~\ref{Fig1a1b} to the right. Because of a smooth interpolation between these two regimes, RRG can be used to describe the period when the matter content of the universe is in the transition from the radiation to NR matter \cite{RRG2018dosReis}.

Friedmann equation for a universe with curvature $\kappa $, cosmological constant $\Lambda $, RRG, NR and UR components can be expressed as follows: 
\begin{equation}
H^{2}=\frac{8\pi G}{3}\left( \rho _{\mathcal{R}}+\rho _{\text{NR}}+\rho _{%
\text{UR}}+\rho _{\Lambda }\right) -\frac{\kappa }{a^{2}}\text{,}
\label{eq FLRW geral 0}
\end{equation}%
where $H\equiv \dot{a}/a$ is the Hubble parameter and "$\cdot $" represent the derivative with respect to time. The simplicity of the equation of state for RRG makes it possible to obtain analytical solutions of Friedmann equation with one or multiple components including RRG (see Refs.~\cite{RRG2005BERREDO-PEIXOTO,RRG2012Leo,RRG2018MNRAS} for details).

For instance, considering only RRG and introducing a new variable $u=a^{2}/a_{\text{ref}}^{2}$ in Eq.~(\ref{eq FLRW geral 0}) we get 
\begin{equation}
\dot{u}^{2}=4\frac{8\pi G}{3}u^{2}\rho _{\mathcal{R}}\left( u\right)
=4H_{0}^{2}\Omega _{d\mathcal{R},\text{ref}}\sqrt{u+b^{2}}\text{,}
\label{Eq13_rrg1}
\end{equation}%
where $\Omega _{d\mathcal{R},\text{ref}}\equiv \rho _{d\mathcal{R},\text{ref}}/\rho _{\text{crit},0}$ with $\rho _{\text{crit},0}=3H_{0}^{2}/8\pi G$. The subscript "$0$" indicates the current value of the quantities. Therefore, integrating Eq.~(\ref{Eq13_rrg1}) and assuming that $a\left( t\simeq 0\right) \simeq 0$, we obtain 
\begin{equation}
a\left( t\right) =a_{\text{ref}}\sqrt{\left( \frac{3H_{0}\sqrt{\Omega _{d%
\mathcal{R},\text{ref}}}}{2}t\text{ }+b^{\frac{3}{2}}\right) ^{\frac{4}{3}%
}-b^{2}\text{.}}  \label{eq a func t e b}
\end{equation}%
Observe that for $a\gg ba_{\text{ref}}$ (NR limit) the standard NR behavior $a\left( t\right) \propto t^{2/3}$ is obtained. On the other hand, assuming $a\ll ba_{\text{ref}}$ (UR limit) leads to typical UR behavior $a\left(t\right) \propto \sqrt{t}$.

\begin{figure}[tbh]
\centering
\includegraphics[scale=1.0]{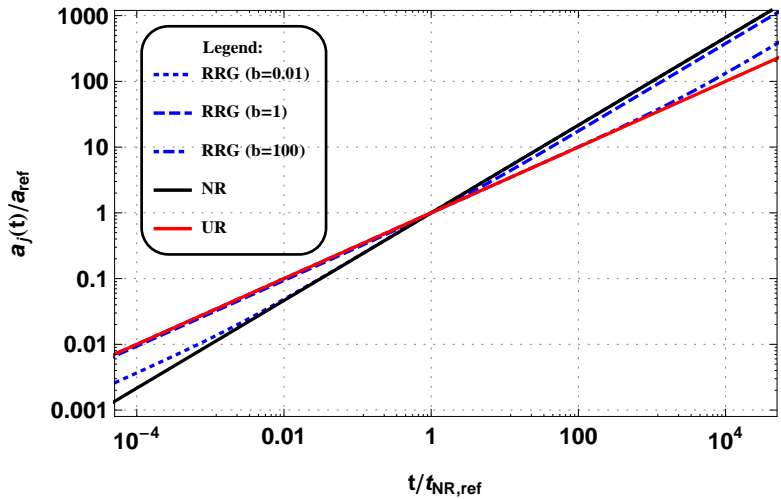}
\caption{Evolution of scale factor with time for a universe where RRG component is dominant. The $t_{\text{NR,ref}}$ is NR time $t_{\text{NR}}=t(b=0)$ in $ a=a_{\text{ref}}$ and the black and red curves serve as the reference since they correspond to NR and UR limits, respectively.}
\label{Fig t x a}
\end{figure}

Therefore, the dependence of the scale factor with the time given by Eq.~(\ref{eq a func t e b}), interpolates between the typical solutions of a universe dominated by UR matter to NR ones at $a/a_{\text{ref}}\sim b$. This statement is illustrated in Fig.~\ref{Fig t x a}, where we plot the Eq.~(\ref{eq a func t e b}) for different values of $b$.

\subsection{The relationship between the parameter \textit{b} and the mass \textit{m}}

In the description via RRG, the relativistic degree is defined by the ratio $a/(a_{\text{ref}}b)$. On the other hand, in the description via Maxwell-Boltzmann statistics the relativistic degree is determined by the ratio of the mass of the particle to its temperature $m/T$. To obtain a consistent relationship between these two descriptions it is reasonable to assume that at some specific time $t_{\text{ref}}$ both are equal and then independently evolve. Thus, at  $t_{\text{ref}}$, the two descriptions have the same energy density and numerical density of particles, i.e.,  $\rho_{\mathcal{R}}\left( a_{\text{ref}}\right) =\rho _{\text{MB}}\left( m/T_{\text{ref}}\right) $ and $n_{\mathcal{R}}\left( a_{\text{ref}}\right) =n_{\text{MB}}\left( m/T_{\text{ref}}\right) $. Moreover, since $a_{\text{ref}}$ is an arbitrary scale factor, we adopt for convenience $a_{\text{ref}}=a_{0}$ which represents the current scale factor.

In this case, combining the Eqs.~(\ref{eq Rho RRG fuc de a e b1}) and (\ref{Rho_k1/k2}) at $t=t_{\text{ref}}=t_{0}$, we can obtain the following relation:
\begin{equation}
b^{2}=\left[ \frac{K_{1}\left( x_{0}\right) }{K_{2}\left( x_{0}\right) }+%
\frac{3}{x_{0}}\right] ^{2}-1\text{,}  
\label{b fun m}
\end{equation}%
where $x_{0}\equiv \rho_{d\text{MB}}/P_{\text{MB}}=m/T_{0}^{\text{MB}}$ and $T_{0}^{\text{MB}}$ is the current temperature of the relativistic Maxwell-Boltzmann gas. Therefore, it is necessary to determine the $T_{0}^{\text{MB}}$ to obtain a unique relationship between $b$ and $m$. The difficulty here is the $T_{0}^{\text{MB}}$ generally depends on the cosmological model adopted. Nevertheless, from reasonable assumptions, we can get a relatively general estimate for it.

To obtain the $T_{0}^{\text{MB}}$ in a simple and approximate way we assume that the gas of relativistic particles undergoes an abrupt transition from UR to NR as the universe expands. In the UR case, one has known that $T_{\text{UR}}^{\text{MB}}\propto a^{-1}$. On the other hand, for NR case, it can be shown from the conservation equation that $T_{\text{NR}}^{\text{MB}}\propto a^{-2}$.

The next step is to assume that the abrupt transition from UR to NR occurs at $T_{\text{tran}}^{\text{MB}}=m$ such that 
\begin{equation}
T^{_{\text{MB}}}\propto\left\{ 
\begin{array}{lll}
D/a & \text{for} & T>m \\ 
Q/a^{2} & \text{for} & T\leq m%
\end{array}
\right. \text{,}
\end{equation}
where $D$ and $Q$ are constants to be determined. In addition, we suppose that at some point in the early universe Maxwell-Boltzmann gas was in equilibrium with the photons. Thus, neglecting any extra contribution of photons due to the  pair annihilation processes\footnote{The most common example of such processes occurs shortly before nucleosynthesis when the annihilation of electron-positron pairs causes a slight increase in photon temperature with respect to neutrino temperature.}, the adiabatic expansion ensures $T_{\gamma }=T_{\text{UR}}^{_{\text{MB}}}$. Thus, since $T_{\gamma}=T_{\gamma0}\left( a_{0}/a\right) $ and $T_{\gamma}=T_{\text{UR}}^{_{\text{MB}}}$, we have $D=T_{\gamma0}a_{0}$. Setting $a\left( T_{\text{tran}}^{\text{MB}}\right) =a_{\text{tran}}$, we have $T_{\text{UR}}^{_{\text{MB}}}\left( a_{\text{tran}}\right) =T_{\text{NR}}^{_{\text{MB}}}\left( a_{\text{tran}}\right) $, so that $Q=T_{\gamma0}a_{0}a_{\text{tran}}$ where $a_{\text{tran}}/a_{0}=T_{\gamma0}/T_{\text{tran}}^{\text{MB}}=T_{\gamma0}/m$. Therefore,%
\begin{equation}
T^{_{\text{MB}}}\left( a\right) =\left\{ 
\begin{array}{lll}
T_{\gamma0}\left( \frac{a_{0}}{a}\right) & \text{for} & T>m \\ 
T_{\gamma0}\left( \frac{T_{\gamma0}}{m}\right) \left( \frac{a_{0}}{a}\right)
^{2} & \text{for} & T\leq m%
\end{array}
\right. \text{.}  \label{eq TMB}
\end{equation}

According to the Eq.~(\ref{eq TMB}), the temperature $T^{_{\text{MB}}}(a) $ calculated today is $T_{0}^{_{\text{MB}}}=( T_{\gamma 0})^{2}/m $, then the parameter $x_{0}$ is given by: 
\begin{equation}
x_{0}=\frac{m}{T_{0}^{_{\text{MB}}}}=\left( \frac{m}{T_{\gamma0}}\right) ^{2}%
\text{.}  \label{x0 modelado}
\end{equation}
Substituting this last expression into Eq.~(\ref{b fun m}), we can obtain an expression for $b$ as a function of mass $m$, where $T_{\gamma0}=2.3486 \times10^{-7}$keV \cite{T0_Fixsen2009}.

In the specific case of WDM, it is natural to expect $m\gg T_{\gamma0}$. Thus, adopting the RRG to describe the WDM, we can use the asymptotic expansions for the Bessel functions to approximate the Eq.~(\ref{b fun m}) obtaining the expression $b^{2}\simeq3/x_{0}$. Finally, using Eq.~(\ref{x0 modelado}) in the latter result, we conclude that%
\begin{equation}
b^{2}\simeq3\left( \frac{T_{\gamma0}}{m}\right) ^{2}\text{ or }m\simeq \sqrt{%
3}\left( \frac{T_{\gamma0}}{b}\right) \simeq\frac{4.07}{10^{7}b}\text{ keV.}
\label{massa b WDM}
\end{equation}
Note that due to the simplifications used this result should be regarded as an approximate relation between $m$ and $b$.

A more complex procedure involving the transfer function of thermal relics \cite{Bode_2001} can be used to obtain an equation analogous to Eq.~(\ref{massa b WDM}). From a numerical method using RRG and thermal relics, W. S. Hip\'{o}lito-Ricaldi \textit{et al.} \cite{RRG2018Hipolito} obtained the following fit-formula: 
\begin{equation}
m_{\omega}=\frac{4.65}{10^{6}b^{4/5}}\text{ keV.}  
\label{m por hipolito}
\end{equation}

\begin{figure}[tbh]
\centering
\includegraphics[scale=1.0]{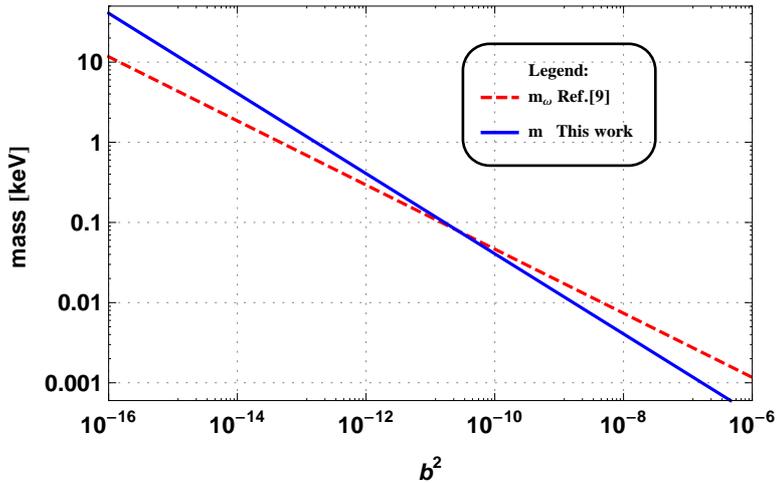}
\caption{Plots of the mass as function of the parameter $b^{2}$ given by Eq.~(\protect\ref{massa b WDM}) (continuous blue curve) and Eq.~(\protect\ref{m por hipolito}) (dashed red curve).}
\label{rel_entre_b_x_m}
\end{figure}

In Fig.~\ref{rel_entre_b_x_m} we plot the mass as a function of the parameter $b^{2}$ given by Eq.~(\ref{massa b WDM}) and Eq.~(\ref{m por hipolito}). Analyzing this figure, we verify that the expression obtained in our work and that obtained in Ref.~\cite{RRG2018Hipolito} are equivalent in order of magnitude. In addition, we observed the largest discrepancy occurs for smaller values of $b$ when we consider a range between $10^{-16}$ and $10^{-6}$. These same conclusions can be obtained from Table~\ref{tab:rel_b_massa} where we assign six values for $b^{2}$ and show the equivalent value of mass and thermal velocity. It is worth noting that constraints from high redshift Lyman-$\alpha$ forest data impose a lower bound $m_{\text{WDM}}\geq3.3$ keV for particles of WDM in the form of early decoupled thermal relics \cite{mWDMMatteoViel2013}. Thus, using Eq.~(\ref{massa b WDM}) and Eq.~(\ref{m por hipolito}) we find that the values of $b^{2}$ corresponding to this mass limit are $b^{2}\lesssim1.52\times10^{-14}$ and $b^{2}\lesssim2.35\times10^{-15}$, respectively. As shown in Ref.~\cite{RRG2018Hipolito}, the RRG reproduces the relative standard WDM transfer function with an accuracy of $1\%$ for $b^{2}\lesssim10^{-14}$.
\begin{table}[th]
\centering
%\rowcolors{1}{green}{pink}%
\begin{tabular}{cccc}
\hline\hline
$\#$ & Ref.~\cite{RRG2018Hipolito} & \multicolumn{2}{c}{This work} \\ 
$b^{2}$ & $m_{\omega}\text{ [keV]}$ & $m\text{ [keV]}$ & $v_{th\mathcal{R},0}%
\text{ [m/s]}$ \\ \hline
\rowcolor{Gray} $10^{-16}$ & 11.68 & 40.7 & 3.00 \\ 
$10^{-15}$ & 4.65 & 12.87 & 9.48 \\ 
\rowcolor{Gray} $10^{-14}$ & 1.85 & 4.07 & 30.00 \\ 
$10^{-13}$ & 0.73 & 1.28 & 94.86 \\ 
\rowcolor{Gray} $10^{-12}$ & 0.29 & 0.40 & 300.00 \\ 
$10^{-11}$ & 0.11 & 0.12 & 948.68 \\ \hline\hline
\end{tabular}%
\caption{Six values for the parameter $b^{2}$ in a scenario of WDM and their corresponding values of mass and thermal velocity.}
\label{tab:rel_b_massa}
\end{table}

\section{Linear perturbation level: distribution function and dynamic equations}\label{perturbation}

In the linear level, we restrict the analyses to scalar perturbations in a spatially flat universe. The scalar metric in the Newtonian gauge is given by%
\begin{equation}
ds^{2}=-\left( 1+2\Psi \right) dt^{2}+a^{2}\left( 1+2\Phi \right) \delta
_{ij}dx^{i}dx^{j}\text{,}  \label{metrica}
\end{equation}%
where the perturbation of the metric $\Psi (x,t)$ corresponds to the Newtonian potential and $\Phi (x,t)$ is associated with the perturbation of the spatial curvature. We define the quadri-momentum as $P^{\nu }\equiv dx^{\nu }/d\lambda $ and the energy by $E\equiv \sqrt{p^{2}+m^{2}}$, where $ m $ and $p^{2}\equiv g_{ij}P^{i}P^{j}$ are the mass and proper momentum of the particle, respectively. Using $g_{\mu \nu }P^{\mu }P^{\nu }=-m^{2}$, it
is possible to show that \cite{Dodelson}:%
\begin{equation}
P^{\mu }=\left[ E\left( 1-\Psi \right) ,p\frac{\left( 1-\Phi \right) }{a}%
\widehat{p}^{i}\right] \text{.}  \label{momento comovel}
\end{equation}%

\subsection{Distribution function and energy-momentum tensor of RRG}

In the linear perturbation level, we relax the basic assumption of RRG by allowing a small variation around the momentum $\bar{p}_{\mathcal{R}}\left( t\right) $ of the particles, so the new momentum will be 
\begin{equation}
p_{\mathcal{R}}\left( \vec{x},\widehat{p},t\right) =\bar{p}_{\mathcal{R}}\left( t\right)
+\delta p_{\mathcal{R}}\left( \vec{x},\widehat{p},t\right) ,
\end{equation}
with $\bar{p}_{\mathcal{R}}\left( t\right) \gg\delta p_{\mathcal{R}}\left( \vec{x},\widehat {p},t\right) $. From here on, our bar will be used to identify background quantities. Observe the momentum $p_{\mathcal{R}}$ is a parameter of distribution function (such as the temperature for photons) which in the background is only time dependent. On the other hand, at a linear perturbation level, we consider the most general dependence for $p_{\mathcal{R}}$, namely, a dependence on $\vec{x}$ (inhomogeneities), $\widehat{p}$ (anisotropies) and obviously on time $t$. By definition of RRG all the particles have the same magnitude of momentum hence we do not consider $p_{\mathcal{R}}$ depending on the magnitude of the moment $p$.

Thus, the RRG distribution function can be rewritten as 
\begin{equation}
f_{\mathcal{R}}\left( \vec{x},p,\widehat{p},t\right) =Cp\delta (p-p_{%
\mathcal{R}})\text{,}  \label{f}
\end{equation}%
where $p_{\mathcal{R}}$ can be interpreted as a constraint parameter of the momentum variable $p$. In linear perturbation level, Eq.~(\ref{f}) can be expanded as 
\begin{equation}
f_{\mathcal{R}}\simeq \bar{f}_{\mathcal{R}}+\left. \frac{\partial f_{%
\mathcal{R}}}{\partial p_{\mathcal{R}}}\right\vert _{p_{\mathcal{R}}=\bar{p}_{\mathcal{R}}%
}\delta p_{\mathcal{R}}=\bar{f}_{\mathcal{R}}+\mathcal{F}_{\mathcal{R}} \text{,}
\end{equation}%
where $\bar{f}_{\mathcal{R}}\left( p,t\right)\equiv Cp\delta \left( p-\bar{p}_{\mathcal{R}}\right) $ is the background RRG distribution function and 
\begin{equation}
\mathcal{F}_{\mathcal{R}}\left( \vec{x},p,\widehat{p},t\right) \equiv -p%
\frac{\partial \bar{f}_{\mathcal{R}} }{\partial p}%
\mathcal{M}\left( \vec{x},\widehat{p},t\right)\equiv -\frac{p}{\bar{p}_{\mathcal{R}}}\frac{\partial \bar{f}_{%
\mathcal{R}}}{\partial p}\delta p_{\mathcal{R}} 
\label{Func de distrib do RRG exp}
\end{equation}%
is the first-order RRG distribution function. Note that, $p_{\mathcal{R}}$ is for RRG distribution function which the temperature $T$ is for the photon distribution function. In this sense, $\bar{p}_{\mathcal{R}}$, $ \delta p_{\mathcal{R}}$ and $\mathcal{M}$ play the role of $\bar{T}$, $\Delta T$ and $ \Theta \equiv \Delta T/\bar{T}$, respectively.

Once the RRG distribution function is established, we can find its energy-momentum tensor from the general expression \cite{Dodelson, Piattella:2018}, 
\begin{equation}
T^{\mu }{}_{\nu }=\int \frac{dP_{1}dP_{2}dP_{3}}{\left( 2\pi \right) ^{3}}%
\frac{1}{\sqrt{-\det \left[ g_{\alpha \beta }\right] }}\frac{P^{\mu }P_{\nu }%
}{P^{0}}f\text{.}  \label{Tensor_Ener_mon_f}
\end{equation}%
To proceed with the calculations, we performed a multipolar expansion of $\mathcal{M}\left( \vec{x},\widehat{p},t\right) $ in Fourier space,%
\begin{equation}
\mathcal{M}\left( k,\mu ,t\right) =\sum_{l=0}^{\infty }\left( -i\right)
^{l}\left( 2l+1\right) \mathcal{M}_{l}\left( k,t\right) \mathcal{P}%
_{l}\left( \mu \right) \text{,}
\end{equation}%
where $P_{l}\left( \mu \right) $ are the Legendre polynomials, $\mu \equiv \widehat{k}^{i}\widehat{p}_{i}=\widehat{k}\cdot \widehat{p}$ and $\mathcal{M}_{l}$ (with $l=0,1,2,\cdots $) are the multipoles of the momentum contrast, given by the expression 
\begin{equation}
\mathcal{M}_{l}\left( k,t\right) =\frac{1}{\left( -1\right) ^{l}}%
\int_{-1}^{1}\frac{d\mu }{2}P_{l}\left( \mu \right) \mathcal{M}\left( k,\mu
,t\right) \text{.}
\end{equation}

Using the metric (\ref{metrica}), the momentum $P^{\mu }$ and the distribution function, we can show that 
\begin{equation}
T_{\text{ \ }\nu }^{\mu }=\bar{T}_{\text{ \ }\nu }^{\mu }+\delta T_{\text{
\ }\nu }^{\mu }=\left[ 
\begin{array}{cc}
-\left( \bar{\rho}_{\mathcal{R}}+\delta \rho _{\mathcal{R}}\right)  & 
-\left( \bar{\rho}_{\mathcal{R}}+\bar{P}_{\mathcal{R}}\right) \frac{v_{%
\mathcal{R}}}{a}\widehat{k}^{i} \\ 
\left( \bar{\rho}_{\mathcal{R}}+\bar{P}_{\mathcal{R}}\right) av_{%
\mathcal{R}}\widehat{k}_{j} & \left( \bar{P}_{\mathcal{R}}+\delta P_{%
\mathcal{R}}\right) \delta ^{i}{}_{j}+\pi _{\mathcal{R}}{}^{i}{}_{j}%
\end{array}%
\right]\text{,}  \label{Eq Tnumu RRG completo}
\end{equation}%
where%
\begin{equation}
\delta \rho _{\mathcal{R}}=-\delta T_{\text{ \ }0}^{0}=3\bar{\rho}_{%
\mathcal{R}}\left( 1+\omega _{\mathcal{R}}\right) \mathcal{M}_{0}
\label{deltarho}
\end{equation}%
and%
\begin{equation}
\delta P_{\mathcal{R}}=\frac{1}{3}\delta ^{j}{}_{i}{}\delta T_{\text{ \ }%
j}^{i}=\bar{P}_{\mathcal{R}}\left( 5-3\omega _{\mathcal{R}}\right) 
\mathcal{M}_{0}  \label{deltaP}
\end{equation}%
are the energy density and pressure perturbations, respectively. The irrotational velocity perturbation $v_{\mathcal{R}}$ in the Fourier space is given by the expression: 
\begin{equation}
v_{\mathcal{R}}=\widehat{k}^{i}v_{\mathcal{R}}{}_{i}\equiv \widehat{k}^{i}%
\frac{\delta T_{\text{ \ }i}^{0}}{\left( \bar{\rho}_{\mathcal{R}}+\bar{P}%
_{\mathcal{R}}\right) a}=-i\frac{4\sqrt{3\omega _{\mathcal{R}}}}{\left(
1+\omega _{\mathcal{R}}\right) }\mathcal{M}_{1}\text{.}  
\label{pertubV}
\end{equation}%
The $\pi _{\mathcal{R}}{}^{i}{}_{j}$ is the traceless longitudinal part of the energy-momentum tensor, defined as%
\begin{equation}
\pi _{\mathcal{R}}{}^{i}{}_{j}\equiv \delta T_{\text{ \ }j}^{i}-\frac{\delta
^{l}{}_{k}}{3}\delta T_{\text{ \ }l}^{k}\delta ^{i}{}_{j}\text{,}
\end{equation}%
and quantified by the variable $\sigma _{\mathcal{R}}$%
\begin{equation}
\sigma _{\mathcal{R}}\equiv \widehat{k}_{i}\widehat{k}^{j}\frac{\pi _{%
\mathcal{R}}{}^{i}{}_{j}}{\left( \bar{\rho}_{\mathcal{R}}+\bar{P}_{%
\mathcal{R}}\right) }=-2\omega _{\mathcal{R}}\left( \frac{5-3\omega _{%
\mathcal{R}}}{1+\omega _{\mathcal{R}}}\right) \mathcal{M}_{2}\text{.}
\label{sigma_R}
\end{equation}%
where $\widehat{k}^{i}\equiv k^{i}/\sqrt{k_{j}k^{j}}$ is the unit wavevector which arises when we take the Fourier transform.

Combining Eq.~(\ref{deltarho}) with Eq.~(\ref{deltaP}), the pressure perturbation is given by 
\begin{equation}
\delta P_{\mathcal{R}}=\frac{\omega _{\mathcal{R}}}{3}\frac{\left( 5-3\omega
_{\mathcal{R}}\right) }{\left( 1+\omega _{\mathcal{R}}\right) }\delta \rho _{%
\mathcal{R}}\text{.}  
\label{relacao de deltaP com deltaRho}
\end{equation}%
This equation can also be obtained perturbing Eq.~(\ref{EoS RRG}) with respect to the momentum by taking into account the role of $\rho _{d\mathcal{R}}$ without referring to any particular gauge. Thus, it can be identified as the sound speed squared in the rest frame of RRG: 
\begin{equation}
c^2_s = \frac{\omega _{\mathcal{R}}}{3}\frac{\left( 5-3\omega
_{\mathcal{R}}\right) }{\left( 1+\omega _{\mathcal{R}}\right) }\,.
\end{equation}
Moreover, one can check $c^2_s$ is equal to the adiabatic sound speed squared, $c^2_a=\dot{P_{\mathcal{R}}}/\dot{\rho_{\mathcal{R}}}$, hence RRG induces no intrinsic entropy perturbations. 

Applying UR limit $\bar{\rho}_{d\mathcal{R}}^{2}/\bar{\rho}_{\mathcal{R}}^{2}\ll 1$ in the previous equations, we obtain 
\begin{equation}
\delta \rho _{\mathcal{R}}=4\bar{\rho}_{\mathcal{R}}\mathcal{M}_{0}\text{,
\ \ \ }v_{\mathcal{R}}=-3i\mathcal{M}_{1}\text{, \ \ \ }\delta P_{\mathcal{R}%
}=\frac{4}{3}\bar{\rho}_{\mathcal{R}}\mathcal{M}_{0}
\end{equation}%
and%
\begin{equation}
\widehat{k}_{i}\widehat{k}^{j}\pi _{\mathcal{R}}{}^{i}{}_{j}=\left( \widehat{%
k}_{i}\widehat{k}^{j}-\frac{\delta ^{j}{}_{i}}{3}\right) \delta T_{\text{ \ }%
j}^{i}=-\frac{8}{3}\bar{\rho}_{\mathcal{R}}\mathcal{M}_{2}\text{.}
\label{projecao PI}
\end{equation}%
Note that these expressions have the same functional form as those obtained for photons and massless neutrinos. On the other hand, applying NR limit $\bar{\rho}_{d\mathcal{R}}^{2}/\bar{\rho}_{\mathcal{R}}^{2}\rightarrow 1$, we obtain similar expressions for those of cold dark matter and baryons:%
\begin{equation}
\delta \rho _{\mathcal{R}}=3\bar{\rho}_{\mathcal{R}}\mathcal{M}_{0}\text{,
\ \ \ }v_{\mathcal{R}}=-i4v_{\text{th}\mathcal{R}}\mathcal{M}_{1}\text{, \ \
\ }\delta P_{\mathcal{R}}=0
\end{equation}%
and%
\begin{equation}
\widehat{k}_{i}\widehat{k}^{j}\pi _{\mathcal{R}}{}^{i}{}_{j}=\left( \widehat{%
k}_{i}\widehat{k}^{j}-\frac{\delta ^{j}{}_{i}}{3}\right) \delta T_{\text{ \ }%
j}^{i}=0\text{.}
\end{equation}%
At NR limit, we can not neglect $\sqrt{3\omega _{\mathcal{R}}}$, because the density perturbations themselves induce velocities through the continuity equation \cite{Dodelson}. Therefore, in the linear treatment of RRG at NR limit, we maintain the $\sqrt{3\omega _{\mathcal{R}}}$\ terms and neglect the $\omega _{\mathcal{R}}$\ ones, which are equivalent to terms of first and second order in $\bar{p}_{\mathcal{R}}/\bar{E}$, respectively.

\subsection{Einstein and conservation equations}

Using the metric (\ref{metrica}) and assuming a universe constituted solely (or dominated) by RRG whose energy-momentum tensor is expressed by Eq.~(\ref{Eq Tnumu RRG completo}), the perturbed Einstein equation in Fourier space leads to the following expressions:%
\begin{align}
3a^{2}H\dot{\Phi}-3a^{2}H^{2}\Psi +k^{2}\Phi & =4\pi Ga^{2}\delta \rho _{%
\mathcal{R}}\text{,}  \label{Eq Einstein 00} \\
-ik\left( H\Psi -\dot{\Phi}\right) & =4\pi Ga\bar{\rho}_{\mathcal{R}%
}\left( 1+\omega _{\mathcal{R}}\right) v_{\mathcal{R}}\text{,}
\label{Eq Einstein 0i} \\
k^{2}\left( \Psi +\Phi \right) & =12\pi Ga^{2}\bar{\rho}_{\mathcal{R}%
}\left( 1+\omega _{\mathcal{R}}\right) \sigma _{\mathcal{R}}
\label{Eq Einstein ij com i dif j}
\end{align}%
and%
\begin{equation}
a^{2}\ddot{\Phi}-a^{2}H\left( \dot{\Psi}-3\dot{\Phi}\right) -a^{2}\left( 2%
\frac{\ddot{a}}{a}+H^{2}\right) \Psi +\frac{k^{2}}{3}\left( \Psi +\Phi
\right) =-4\pi Ga^{2}\delta P_{\mathcal{R}}.
\label{Eq Einstein ij com i igual j}
\end{equation}

In this same context, the conservation equation of the energy-momentum tensor leads to the well-known energy and momentum equation for fluids with the correspond RRG quantities:%
\begin{equation}
\dot{\delta}_{\mathcal{R}}
+3H\left(c^2_s - \omega _{\mathcal{R}}\right)\delta_{\mathcal{R}}
+\left(1+ \omega _{\mathcal{R}} \right)\left(\frac{ik}{a}v_{\mathcal{R}}+3\dot{\Phi}\right)=0
\label{Eq conserv 1 via T_numu}
\end{equation}
and%
\begin{equation}
\dot{v}_\mathcal{R} + H\left( 1 -3c^2_s\right)v_{\mathcal{R}} 
+\frac{ik}{a}\left( \frac{c^2_s}{1+\omega_\mathcal{R} }\delta _\mathcal{R} + \sigma_{\mathcal{R}}+\Psi \right)=0\,,
\label{Eq conserv 2 via T_numu}
\end{equation}
where $\delta_{\mathcal{R}}=\delta \rho _{\mathcal{R}}/\bar{\rho} _{\mathcal{R}}$ is the density contrast. As expected, in the absence of interactions, we obtain equations with functional forms identical to those of cold dark matter (or baryons) and massless neutrinos (or photons), when we apply NR and UR limits, respectively.

It is important to emphasize that up to now, our description of RRG in the perturbed cosmological context does not constitute a closed system of the equation since there are six variables ($\Phi,\Psi,\delta\rho_{\mathcal{R}},v_{\mathcal{R}},\sigma_{\mathcal{R}},\delta P_{\mathcal{R}}$) and only five independent equations namely Eqs.~(\ref{Eq Einstein 00}), (\ref{Eq Einstein 0i}), (\ref{Eq Einstein ij com i dif j}), (\ref{Eq Einstein ij com i igual j}) and (\ref{relacao de deltaP com deltaRho}). If for some physical reason the anisotropic stress term $\sigma _{\mathcal{R}}$ can be neglected or determined as a function of other variables (e.g. Ref.~\cite{Shoji:2010hm}) this set of equations becomes a closed set. Otherwise, it is necessary to use the Boltzmann equation to complete the description. In the following section, we study RRG in this context.

\subsection{Boltzmann equation}

The Boltzmann equation in the differential form is given by 
\begin{equation}
\frac{df\left(\vec{x},p,\widehat{p},t\right)}{dt}=C[f\left(\vec{x},p,\widehat{p},t\right)].  \label{Eq-Bolt}
\end{equation}
The right-hand side represents the collision terms $C[f]$ which are usually complicated functions of the distribution function. In the Newtonian gauge, we can express the complete derivative $df/dt$ in terms of partial derivatives, so 
\begin{equation}
\frac{df}{dt}=\frac{\partial f}{\partial t}+\frac{\partial f}{\partial x^{i}}%
\frac{p}{E}\frac{\widehat{p}^{i}}{a}-\frac{E}{p}\frac{\partial f}{\partial p}%
\left[ \frac{p}{a}\frac{\partial\Psi}{\partial x^{j}}\widehat{p}^{j}+\frac{%
p^{2}}{E}\left( H+\frac{\partial\Phi}{\partial t}\right) \right] =C[f].
\label{Eq-Bolt FD generica}
\end{equation}

For the zero order of perturbation, the previous equation becomes: 
\begin{equation}
\left. \frac{df}{dt}\right\vert _{\text{zero order}}=\frac{\partial\bar{f}%
}{\partial t}-Hp\frac{\partial\bar{f}}{\partial p}=0,  \label{eq_Bolt_0}
\end{equation}
where we establish $C[f]=0$ since we are interested only in the collisionless Boltzmann equation. In the specific case of RRG, we can manipulate the previous expression such that 
\begin{equation}
\frac{\partial}{\partial p}\left[ \left( \frac{p}{\bar{p}_{\mathcal{R}}}\frac{d\bar{p}_{\mathcal{R}}}{dt}%
\frac{\partial\bar{f}_{\mathcal{R}}}{\partial p}+Hp\right) \bar{f}_{%
\mathcal{R}}\right] +\frac{2}{\bar{p}_{\mathcal{R}}}\frac{d\bar{p}_{\mathcal{R}}}{dt}\bar{f}_{\mathcal{%
R}}=-2H\bar{f}_{\mathcal{R}}\text{.}
\end{equation}
Hence, integrating both sides with respect $p$, we conclude that%
\begin{equation}
-\frac{1}{\bar{p}_{\mathcal{R}}}\frac{d\bar{p}_{\mathcal{R}}}{dt}=H\Rightarrow\bar{p}_{\mathcal{R}}\propto\frac{1}{a}.
\end{equation}
This if implies the RRG distribution function satisfies the zero-order Boltzmann equation $\bar{p}_{\mathcal{R}}\propto a^{-1}$. This is expected since we can show from the geodesic equation that the momentum $p$ of any particle is proportional to $a^{-1}$ in a Friedmann universe.

For the first-order of perturbation, we can express the left-hand side of Eq.~(\ref{Eq-Bolt FD generica}) as 
\begin{equation}
\left. \frac{df_{\mathcal{R}}}{dt}\right\vert _{\text{first-order}}=\frac{%
\partial \mathcal{F}_{\mathcal{R}}}{\partial t}+\frac{\partial \mathcal{F}_{%
\mathcal{R}}}{\partial x^{i}}\frac{p}{E}\frac{\widehat{p}^{i}}{a}-pH\frac{%
\partial \mathcal{F}_{\mathcal{R}}}{\partial p}-\frac{E}{p}\frac{\partial 
\bar{f}_{\mathcal{R}}}{\partial p}\left( \frac{p}{a}\frac{\partial \Psi }{%
\partial x^{j}}\widehat{p}^{j}+\frac{p^{2}}{E}\frac{\partial \Phi }{\partial
t}\right) \text{.}  \label{Eq Bolt ordem1}
\end{equation}%
Using Eq.~(\ref{Func de distrib do RRG exp}), the relation $d\bar{p}_{\mathcal{R}}/dt=-\bar{p}_{\mathcal{R}} H$ and properties of the Dirac delta function we can rewrite the time derivative in terms of $p$. In the absence of collision terms, we obtain the following expression for the Boltzmann equation in the Fourier space: 
\begin{equation}
\left. \frac{df_{\mathcal{R}}}{dt}\right\vert _{\text{first-order}}=\left( -p%
\frac{\partial \bar{f}_{\mathcal{R}}}{\partial p}\right) \left( \frac{%
\partial \mathcal{M}}{\partial t}+\frac{\partial \Phi }{\partial t}+\frac{ik%
}{a}\mu \frac{p}{E}\mathcal{M}+\frac{ik}{a}\mu \frac{E}{p}\Psi \right) =0.
\label{Eq Bolt ordem 1 sem colisao}
\end{equation}

This expression is directly related to the equation of energy-momentum tensor conservation since by multiplying Eq.~(\ref{Eq Bolt ordem 1 sem colisao}) by $Ed^{3}p/(2\pi)^{3}$ and integrating it, we obtain Eq.~(\ref{Eq conserv 1 via T_numu}). Similarly, multiplying Eq.~(\ref{Eq Bolt ordem 1 sem colisao}) by $p\mu d^{3}p/(2\pi)^{3}$ and integrating it, we obtain Eq.~(\ref{Eq conserv 2 via T_numu}). It is worth noting that the first term between parentheses in Eq.~(\ref{Eq Bolt ordem 1 sem colisao}) can not be neglected, since $\bar{f}_{\mathcal{R}}=Cp\delta\left(p-\bar{p}_{\mathcal{R}}\right) $ and therefore it must be considered at the time of integration.

To get new equations necessary to have a closed system of equations, we must go to higher orders of $\mu $. We know the highest multipolar moments are always subdominant compared to their predecessors, and they decay rapidly when the particles become NR. Thus, it is possible to choose a much smaller $l_{\text{max}}$ for the massive particles than for the approximately massless ones \cite{CosmoPerturSynchNewto1995}. If the interest is to apply RRG to describe slightly relativistic particles, where the quadrupole and higher terms can be neglected, there is no need to use Boltzmann hierarchical equations. However, if we wish to apply RRG to describe relativistic or UR particles, where the term of quadrupole and higher are relevant, it is necessary to use Boltzmann hierarchical equations to complete the system of equations.

The Boltzmann hierarchical equations are obtained multiplying Eq.~(\ref{Eq Bolt ordem 1 sem colisao}) by $p\mu ^{m}d^{3}p/(2\pi )^{3}$ where $m=0,1,2,\cdots $ and integrating it. Proceeding in this way, it is possible to obtain an infinite set of hierarchical equations:%
\begin{align}
\frac{\partial \mathcal{M}_{0}}{\partial t}+\frac{k}{a}\frac{\left(
5-3\omega _{\mathcal{R}}\right) }{4}\sqrt{3\omega _{\mathcal{R}}}\mathcal{M}%
_{1}+\frac{\partial \Phi }{\partial t}& =0\text{ \ \ }\left( l=0\right) , \\
3\sqrt{3\omega _{\mathcal{R}}}\frac{\partial \mathcal{M}_{1}}{\partial t}+%
\frac{k}{a}\frac{3\omega _{\mathcal{R}}\left( 5-3\omega _{\mathcal{R}%
}\right) }{4}\left( 2\mathcal{M}_{2}-\mathcal{M}_{0}\right) -\frac{k}{a}%
\frac{3\left( 1+\omega _{\mathcal{R}}\right) }{4}\Psi & =0\text{ \ \ }\left(
l=1\right) , \\
\left( 2l+1\right) \sqrt{3\omega _{\mathcal{R}}}\frac{\partial \mathcal{M}%
_{l}}{\partial t}+\frac{k}{a}\frac{3\omega _{\mathcal{R}}\left( 5-3\omega _{%
\mathcal{R}}\right) }{4}\left[ \left( l+1\right) \mathcal{M}_{\left(
l+1\right) }-l\mathcal{M}_{\left( l-1\right) }\right] & =0\text{ \ \ }\left(
l\geq 2\right) .
\end{align}%
This set of equations governs the evolution of RRG distribution in the phase space. Note that in UR limit, where $\bar{\rho}_{d\mathcal{R}}^{2}/\bar{\rho}_{\mathcal{R}}^{2}\ll 1$, the Boltzmann hierarchical equations for RRG have the same functional form as the massless neutrinos and the collisionless photons (see, e.g., Ref.~\cite{Piattella:2018}).

The practical use of this infinite set of hierarchical equations occurs through truncation in some maximal multipolar order $l_{\text{max}}$. A simple but imprecise method is to define $\mathcal{M}_{l}=0$ for $l>l_{\text{max}}$. A better truncation scheme is based on the extrapolation of the behavior of $\mathcal{M}_{l}$ to $l=l_{\text{max}}+1$, presented in Ref.~\cite{CosmoPerturSynchNewto1995}.

The deduction of RRG distribution function makes it possible to study its dynamics through the Boltzmann equation, which in turn allows the introduction of collision terms in a natural way. For example, possible applications of RRG include self-interaction and dissipative dark matter, which can solve some small scale discrepancies between predictions of $\Lambda $CDM model and small scale observations
 \cite{Hannestad2000,mWDMMatteoViel2013,Rev_tulin2018,Rocha2013,Peter2013,Zavala2013,Elbert2015,Foot:2016wvj}.

\subsection{Gauge invariant equations}

So far we are working with scalar perturbations in a Newtonian gauge. However, there are other different gauges each one with its own advantages. Thus, the ability to switch between different coordinate systems is useful. To facilitate this process we will rewrite all relevant equations using gauge invariant variables.

In general, the metric scalar perturbation can be written in terms of four quantities \cite{Piattella:2018}%
\begin{equation}
g_{\mu \nu }=a^{2}\left( \eta \right) \times \left[ 
\begin{array}{cc}
-\left( 1+2\Psi \right)  & \partial _{i}w \\ 
\partial _{i}w & \left( 1+2\Phi \right) \delta _{ij}+\left( \partial
_{i}\partial _{j}-\frac{\delta _{ij}}{3}\nabla ^{2}\right) 2\mu 
\end{array}%
\right] \text{,}
\end{equation}%
where $\nabla ^{2}\equiv \delta ^{lm}\partial _{l}\partial _{m}$ and $\eta $ is the conformal time. In addition, with these four quantities, we can construct two invariant gauge variables,%
\begin{equation}
\Psi ^{\left( gi\right) }=\Psi +\frac{1}{a}\left[ \left( w-\mu ^{\prime
}\right) a\right] ^{\prime }\text{ \ \ \ and \ \ \ }\Phi ^{\left( gi\right)
}=\Phi +aH\left( w-\mu ^{\prime }\right) -\frac{1}{3}\nabla ^{2}\mu \text{.}
\end{equation}%
In a similar way, we can construct invariant gauge variables for density, velocity, anisotropic stress and pressure perturbations, so that we get:%
\begin{equation}
\delta \rho ^{\left( gi\right) }=\delta \rho +\bar{\rho}^{\prime }\left(
w-\mu ^{\prime }\right) \text{, \ \ \ }v^{\left( gi\right) }=v-\left( w-\mu
^{\prime }\right) \text{, \ \ \ }\sigma ^{\left( gi\right) }=\sigma 
\end{equation}%
and%
\begin{equation}
\delta P^{\left( gi\right) }=\delta P+\bar{P}^{\prime }\left( w-\mu
^{\prime }\right) \text{,}
\end{equation}%
respectively. Note that, the Newtonian gauge is realized choosing $w=\mu =0$. Thus, the equations obtained in previous sections are easily converted into gauge invariant equations performing the change $X_{N}\rightarrow X^{\left( gi\right) }$.

Therefore, the complete set of gauge invariant equations for RRG, takes the following form:

\textit{Einstein equations}:%
\begin{eqnarray}
3\mathcal{H}\Phi ^{\left( gi\right) }{}^{\prime }-3\mathcal{H}^{2}\Psi
^{\left( gi\right) }+k^{2}\Phi ^{\left( gi\right) } &=&4\pi Ga^{2}\bar{\rho%
}_{\mathcal{R}}\delta _{\mathcal{R}}^{\left( gi\right) }\text{,}
\label{Einstein 00} \\
-ik\left( \mathcal{H}\Psi ^{\left( gi\right) }-\Phi ^{\left( gi\right)
}{}^{\prime }\right) &=&4\pi Ga^{2}\bar{\rho}_{\mathcal{R}}\left( 1+\omega
_{\mathcal{R}}\right) v_{\mathcal{R}}^{\left( gi\right) }\text{,}
\label{Einstein 0i} \\
k^{2}\left( \Psi ^{\left( gi\right) }+\Phi ^{\left( gi\right) }\right)
&=&12\pi Ga^{2}\bar{\rho}_{\mathcal{R}}\left( 1+\omega _{\mathcal{R}%
}\right) \sigma _{\mathcal{R}}^{\left( gi\right) }  \label{Einstein i dif j}
\end{eqnarray}%
and%
\begin{equation}
\Phi ^{\left( gi\right) }{}^{\prime \prime }+2\mathcal{H}\Phi ^{\left(
ig\right) }{}^{\prime }-\mathcal{H}\Psi ^{\left( gi\right) }{}^{\prime
}-\left( 2\mathcal{H}^{\prime }+\mathcal{H}^{2}\right) \Psi ^{\left(
ig\right) }+\frac{k^{2}}{3}\left( \Psi ^{\left( gi\right) }+\Phi ^{\left(
gi\right) }\right) =-4\pi Ga^{2}\delta P_{\mathcal{R}}^{\left( gi\right) }%
\text{.}  \label{Einstein ii}
\end{equation}

\textit{Conservation equations}:%
\begin{equation}
\delta _{\mathcal{R}}^{\left( gi\right) }{}^{\prime }+3\mathcal{H}\left(c_{s}^{2}-\omega _{\mathcal{R}} \right) %
\delta _{\mathcal{R}}^{\left( gi\right) }+\left( 1+\omega _{%
\mathcal{R}}\right) \left(ikv_{\mathcal{R}}^{\left( gi\right)}+ 3\Phi ^{\left( gi\right) }{}^{\prime }  \right)=0
\label{Conserv Boltzmann l 0}
\end{equation}%
and%
\begin{equation}
v_{\mathcal{R}}^{\left( gi\right) }{}^{\prime }+\mathcal{H}\left(1-3c^{2}_{s} \right) v_{\mathcal{R}}^{\left( gi\right) } +ik \left( \frac{c_{s}^{2}}{1+ \omega _{\mathcal{R}}}\delta _{\mathcal{R}}^{\left( gi\right) }+\sigma _{%
\mathcal{R}}^{\left( gi\right) }+\Psi ^{\left(
gi\right) }\right)=0\text{.}  \label{Conserv Boltzmann l 1}
\end{equation}

\textit{Boltzmann hierarchical equation}:%
\begin{align}
\mathcal{M}_{0}^{\prime }+\frac{\left( 5-3\omega _{\mathcal{R}}\right) }{4}%
\sqrt{3\omega _{\mathcal{R}}}k\mathcal{M}_{1}+\Phi ^{\left( gi\right) \prime
}& =0\text{ \ \ }\left( l=0\right) \text{,} \\
3\sqrt{3\omega _{\mathcal{R}}}\mathcal{M}_{1}^{\prime }+3\omega _{\mathcal{R}%
}\frac{\left( 5-3\omega _{\mathcal{R}}\right) }{4}k\left( 2\mathcal{M}_{2}-%
\mathcal{M}_{0}\right) -\frac{3\left( 1+\omega _{\mathcal{R}}\right) }{4}%
k\Psi ^{\left( gi\right) }& =0\text{ \ \ }\left( l=1\right) \text{,} \\
\left( 2l+1\right) \sqrt{3\omega _{\mathcal{R}}}\mathcal{M}_{l}^{\prime
}+3\omega _{\mathcal{R}}\frac{\left( 5-3\omega _{\mathcal{R}}\right) }{4}k%
\left[ \left( l+1\right) \mathcal{M}_{\left( l+1\right) }-l\mathcal{M}%
_{\left( l-1\right) }\right] & =0\text{ \ \ }\left( l\geq 2\right) \text{,}
\label{hierarq Boltzmann l 2}
\end{align}%
where prime indicates derivative with respect to conformal time $\eta$, $\mathcal{H}\equiv aH$ and $\delta _{\mathcal{R}}^{\left( gi\right) }\equiv \delta \rho _{\mathcal{R}}^{\left( gi\right) }/\bar{\rho}_{\mathcal{R}}$.

\section{Applications and discussions}\label{application}

So far we have derived general equations for the cosmological evolution of RRG first-order perturbations. In the following, we consider an application of RRG as a model independent description of WDM and perform a pedagogical application. 

We study the warm approximation of RRG, which can be used to study WDM and, based on current limits for DM mass, argue that, even in the case of very small thermal velocities, RRG is relativistic in the early universe. Then, we derive the initial conditions for RRG perturbations at this epoch. We also study the cosmological evolution of the gravitational potential in a toy model where the universe is solely composed by RRG. 

\subsection{RRG in the warm approximation}

One of the most interesting applications of RRG model is for the description of WDM, which, for instance, can be keV sterile neutrinos \cite{Adhikari_2017}. If dark matter is cold, its simplest representation is a pressureless perfect fluid and the only relevant variables are density and fluid velocity. However, if dark matter has a small but non-negligible thermal velocity, it is important to study the impact of pressure and anisotropic stress. Such studies were conducted using Maxwell-Boltzmann distribution function, e.g., \cite{Armendariz-Picon:2013jej,Piattella:2015nda}. Since the error of RRG model with respect to Maxwell-Boltzmann description for non-relativistic thermal velocities is essentially negligible, a warm fluid can also be studied in this framework. 

Let us consider that the total energy density is given by a small increment above the rest energy density
\begin{equation}
\frac{\rho_{\mathcal{R}}}{\rho_{\mathcal{R}d}}=1+x,\text{ with }0<x\ll 1\, .
\label{rho_expation}
\end{equation}
Using  Eqs.~(\ref{Vth_RRG}) and (\ref{W RRG}) we can expand the fluid quantities, Eqs.~(\ref{deltarho}), (\ref{deltaP}), (\ref{pertubV}) and (\ref{sigma_R}) in terms of $x$:
\begin{equation}
\delta\rho_{\mathcal{R}} / \bar{\rho}_{\mathcal{R}} = \left(3+2x+\mathcal{O}(x^2)\right)\mathcal{M}_{0}\,,
\end{equation}
\begin{equation}
\delta P_{\mathcal{R}} /\bar{\rho}_{\mathcal{R}}   = \left(\frac{10}{3}x-\frac{19}{3}x^{2}+\mathcal{O}(x^3)\right)\mathcal{M}_{0}\, ,
\end{equation}
\begin{equation}
v_{\mathcal{R}}=-i\left(4\sqrt{2x}-\frac{17\sqrt{2}}{3}x^{3/2}+\mathcal{O}(x^{5/2})\right)\mathcal{M}_{1}\,,
\end{equation}
\begin{equation}
\sigma _{\mathcal{R}} = \left(-\frac{20}{3}x+\frac{154}{9}x^{2}+\mathcal{O}(x^3)\right)\mathcal{M}_{2}\,.
\end{equation}
In order to describe a cold fluid we need to retain $x^0$ and $x^{1/2}$ orders, which are associated with density and fluid velocity, respectively. Beyond $x^{1/2}$ order, pressure and anisotropic stress are introduced with $x^{1}$ terms and energy density receives a correction of the same order. This is the first correction to the cold fluid description, which we call warm fluid, i.e., all terms up to $x^1$ order.  

We can relate $x$ with the warmness parameter $b$. Using Eq.~(\ref{W RRG}) we get 
\begin{equation}
\frac{a_{{\rm ref}}}{a}b = \sqrt{2x}+\frac{x^{3/2}}{2\sqrt{2}}+ \mathcal{O}(x^{5/2})\,.
\end{equation}
The sound speed and $\omega_{\mathcal{R}}$ can also be written as a function of $x$:
\begin{equation}
c_{s}^{2}=\frac{10}{9}x-\frac{77}{27}x^{2}+\mathcal{O}(x^{3})\,,
\end{equation} 
\begin{equation}
\omega_\mathcal{R}=\frac{2}{3}x-x^{2} +\mathcal{O}(x^3)\,.
\end{equation}
Then, up to $x^1$ order we have
\begin{equation}
c_{s}=\frac{\sqrt{5}}{3}\frac{a_{{\rm ref}}}{a}b\,.
\end{equation}

Once warmness is introduced, in principle, we need to consider pressure, anisotropic stress and all the higher order multipoles in Boltzmann equations for RRG. Up to order $x^1$, these equations can be written as
\begin{equation}
\mathcal{M}'_{0}+\frac{3\sqrt{5}}{4}c_sk\mathcal{M}_{1}+\Phi'=0\,,
\end{equation}
\begin{equation}
\mathcal{M}'_{1}+\frac{\sqrt{5}}{4}c_sk\left(2\mathcal{M}_{2}-\mathcal{M}_{0}\right)-\frac{\sqrt{5}}{12}\frac{k}{c_s}\Psi=0\,,
\end{equation}
\begin{equation}
\mathcal{M}'_{l}+\frac{\sqrt{5}}{12\left(2l+1\right)}c_sk\left[\left(l+1\right)\mathcal{M}_{l+1}-l\mathcal{M}_{l-1}\right]=0\,.
\end{equation}
Note that we can not take $c_s\rightarrow0$ because this corresponds to neglect the fluid velocity already at cold approximation.

However, the usual argument used for relativistic particles to neglect multipoles with $l\ge2$ on scales larger then horizon, $k\eta\ll1$, can be rephrased for warm RRG in terms of scales larger than the sound horizon $c_s k \eta \ll 1$, which is much smaller then horizon in the warm approximation. Therefore, pressure becomes important only on scales smaller then the sound horizon of the warm fluid, $c_s k\eta >1$ and, as usual, higher multipoles will be important on even smaller scales. This explains the fact that the perfect fluid approximation of  RRG performed in \cite{RRG2018Hipolito} can reproduce with very good accuracy the transfer function for relics computed with the full set of Einstein-Boltzmann equations \cite{Bode_2001}. For instance, considering $b^2 \sim 10^{-14}$ ($a_{\rm ref}=1$ assumed), which corresponds to particles with $m \sim \text{keV}$, pressure becomes important for scales roughly $k>10^3$ $\text{h}/\text{Mpc}$ at $z=0$. Assuming matter domination ($a\propto \eta^2$)  from now until matter-radiation equality, $a_{\rm eq}=10^{-4}$, the scale at which pressure and higher multipoles become important increases to $k>10$ $\text{h}/\text{Mpc}$ until matter-radiation equality. During radiation dominated era ($a\propto\eta$), this scale remains unchanged as long as the warm approximation is valid.

Naturally, at even earlier times, the warm approximation breaks down and higher multipoles have to be considered. From Eq.~(\ref{b fun m}) in the non-relativistic limit, we have $b^2 \simeq 3(T_0/m)$, then, considering only the non-relativistic evolution of temperature, $T\propto a^{-2}$, the temperature and mass will be roughly the same at $a\sim b$, which can also be seen in Fig.~\ref{Fig1a1b}. Hence, keV particles, which corresponds to $b^2 = 10^{-14}$, become relativistic at $a \sim 10^{-7}$ (curiously at this time the temperature of photons is roughly the same). In this case, the warm approximation breaks down very deep inside the radiation dominated era. Therefore, even WDM models that are very non-relativistic today are relativistic in the early universe. In the following, we consider this scenario in order to determine the initial conditions of RRG perturbations.       

\subsection{Initial conditions assuming adiabatic primordial modes}

Let us consider adiabatic initial conditions for RRG perturbations in the early universe. As we saw, if we use RRG to describe particles with $m \simeq$ keV, they will be relativistic for $a<10^{-7}$. Typically, one is interested to determine the evolution of cosmological perturbations after neutrino decoupling, which corresponds to universe with $T<1$ MeV, or $a > 10^{-10}$. In this scenario, we can safely consider RRG as a relativistic component. According to what was discussed in the previous section, this would not be case only for $b^2\lesssim 10^{-20}$, from Eq.~(\ref{massa b WDM}) this corresponds to particles with $m\gtrsim 407$ keV. For simplicity, hereafter we drop the use of the superscript "$\left(gi\right) $".

In the described situation, RRG and neutrinos (see, e.g., Ref.~\cite{Piattella:2018}) will have analogous initial conditions. Therefore, in a super-horizon regime, we have 
\begin{equation}
\frac{\delta _{\mathcal{R}}}{4}=-\Phi +\zeta \text{, \ \ \ }v_{\mathcal{R}%
}=-i\frac{\left( k\eta \right) }{2}\Psi
\end{equation}%
and 
\begin{equation}
\mathcal{M}_{l}=\frac{k\eta }{2l+1}\mathcal{M}_{l-1}\text{ \ for \ }l\geq 2%
\text{,}
\end{equation}%
where $\zeta $ is the comoving curvature perturbation. In addition, the initial value for $\Phi $ and $\Psi $ is given by%
\begin{equation}
\Phi =2\zeta \left[ \frac{5+2R}{15+4R}\right] \text{ \ \ \ and \ \ \ }\Psi =-%
\frac{10\zeta }{15+4R}\text{,}
\end{equation}%
with 
\begin{equation}
R\equiv \sum R_{j}=\frac{1}{\bar{\rho}_{\text{tot}}}\sum \bar{\rho}_{j}%
\text{,}
\end{equation}%
where $\sum \bar{\rho}_{j}$ represents the sum of all non-interacting relativistic components and $\bar{\rho}_{\text{tot}}\simeq \bar{\rho}_{\gamma }+\sum \bar{\rho}_{j}$ in an era dominated by relativistic components.

For example, if one wants to describe standard model neutrinos with RRG  in a universe with cold dark matter, then we have%
\begin{equation}
R=R_{\mathcal{R}}=R_{\nu}\approx\frac{N_{\nu}^{\text{eff}}\left( 7/8\right)
(4/11)^{4/3}}{1+N_{\nu}^{\text{eff}}\left( 7/8\right) (4/11)^{4/3}}\text{,}
\end{equation}
where $\bar{\rho}_{\text{tot}}\simeq\bar{\rho}_{\gamma}+\bar{\rho }_{\mathcal{R}}$ and $N_{\nu}^{\text{eff}}\equiv N_{\nu}g_{\nu}=3.0395$ \cite{MANGANO20028} is the effective number of families. On the other hand, for the description of WDM with RRG we have%
\begin{equation}
R=R_{\mathcal{R}}+R_{\nu}\approx\frac{\frac{\bar{\rho}_{\mathcal{R}}}{%
\rho_{\gamma}}+N_{\nu}^{\text{eff}}\left( 7/8\right) (4/11)^{4/3}}{1+\frac{%
\bar{\rho}_{\mathcal{R}}}{\bar{\rho}_{\gamma}}+N_{\nu}^{\text{eff}%
}\left( 7/8\right) (4/11)^{4/3}}\text{,}
\end{equation}
where $\bar{\rho}_{\text{tot}}\simeq\bar{\rho}_{\gamma}+\bar{\rho}_{\nu}+\bar{\rho}_{\mathcal{R}}$ and $\bar{\rho}_{\mathcal{R}}/\bar{\rho}_{\gamma}$ is the initial energy density fraction of WDM relative to photons. In this case, the quantity $\bar{\rho}_{\mathcal{R}}/\bar{\rho}_{\gamma}$ contributes as a new relativistic species which can be determined by observational data.

\subsection{Potential evolution}

As discussed previously, RRG interpolates between a typical behavior of a UR component to a typically NR component (see Fig.~\ref{Fig1a1b}). In this way, the RRG can be used to approximately describe the background where the relevant constituents are only UR and NR matter \cite{RRG2005BERREDO-PEIXOTO,RRG2018dosReis}. In a similar context, we analyze the evolution of the gravitational potential $\Phi \left(k,a\right) $ in the linear level of perturbation, assuming a hypothetical universe constituted only by RRG with negligible anisotropic stress.

Neglecting the anisotropic stress ($\sigma_{\mathcal{R}}\simeq0$), Eq.~(\ref{Einstein i dif j}) leads to $\Phi=-\Psi$. Thus, combining Eqs.~(\ref{Einstein 00}) and (\ref{Einstein ii}), we obtain a closed-form equation for the potential,%
\begin{equation}
\Phi^{\prime\prime}+3\mathcal{H}\left[ 1+\omega_{\mathcal{R}}\frac{\left(
5-3\omega_{\mathcal{R}}\right) }{\left( 3+3\omega_{\mathcal{R}}\right) }%
\right] \Phi^{\prime}=\left[ 3\omega_{\mathcal{R}}\mathcal{H}^{2}-\left( 3%
\mathcal{H}^{2}+k^{2}\right) \omega_{\mathcal{R}}\frac{\left( 5-3\omega_{%
\mathcal{R}}\right) }{\left( 3+3\omega_{\mathcal{R}}\right) }\right] \Phi%
\text{,}  \label{eq phi sem quadru}
\end{equation}
where $\mathcal{H}$ can be obtained by manipulating Eqs.~(\ref{eq FLRW geral 0}) and (\ref{eq Rho RRG fuc de a e b1}).

Applying the limit $a/a_{\text{ref}}\ll b$ in Eq.~(\ref{eq phi sem quadru}), we obtain%
\begin{equation}
\Phi ^{\prime \prime }+\frac{4}{\eta }\Phi ^{\prime }+\frac{k^{2}}{3}\Phi =0%
\text{,}  \label{eq phi 1}
\end{equation}%
whose solution is \cite{Dodelson} 
\begin{equation}
\Phi =3\Phi _{\text{p}}\left[ \frac{\sin \left( k\eta /\sqrt{3}\right)
-\left( k\eta /\sqrt{3}\right) \cos \left( k\eta /\sqrt{3}\right) }{\left(
k\eta /\sqrt{3}\right) ^{3}}\right] \text{.}  \label{solu phi}
\end{equation}%
This solution shows that in a super-horizon regime $k\eta \ll 1$, the potential $\Phi $ is constant, equal to its primordial value $\Phi_{\text{p}}$. However, when the modes enter the horizon $k\eta >1$, $\Phi $ amplitude rapidly decays with $\left( k\eta \right) ^{-2}$.

Otherwise, applying the limit $a/a_{\text{ref}}\gg b$ in Eq.~(\ref{eq phi sem quadru}), we obtain%
\begin{equation}
\Phi ^{\prime \prime }+\frac{6}{\eta }\Phi ^{\prime }=0\text{,}
\label{eq phi 2}
\end{equation}%
whose general solution is $\Phi =C_{1}+C_{2}\left( k\eta \right) ^{-5}$. Neglecting the decay mode, this solution shows that the gravitational potential is constant when $a/a_{\text{ref}}\gg b$. Therefore, we find that Eq.~(\ref{eq phi 1}) and Eq.~(\ref{eq phi 2}), as well as their solutions, are functionally the same as the expressions obtained for the gravitational potential $\Phi $ in an era dominated by the radiation and dust, respectively.

\begin{figure}[tbh]
\centering
\includegraphics[scale=1.0]{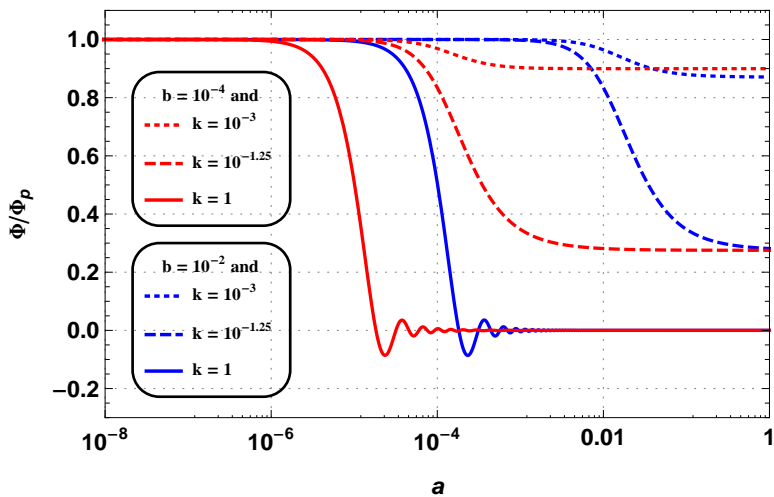}
\caption{Numerical solutions for the linear evolution of gravitational potential $\Phi\left(k,a\right) $, assuming $H_{0}=70$ (km/s)/Mpc and different values for parameter $b$ and $k$ [Mpc$^{-1}$].}
\label{figPhibk}
\end{figure}

In order to obtain a complete description of the evolution of the potential, we define $a_{\text{ref}}=a_{0}\equiv 1$ and perform a change of variable from $\eta $ to $a$ in Eq.~(\ref{eq phi sem quadru}). Thus, we obtain%
\begin{equation}
\frac{d^{2}\Phi }{da^{2}}+\frac{1}{a}\left[ \frac{7}{2}+3\omega _{\mathcal{R}%
}\left( \frac{5-3\omega _{\mathcal{R}}}{3+3\omega _{\mathcal{R}}}-\frac{1}{2}%
\right) \right] \frac{d\Phi }{da}=\frac{\omega _{\mathcal{R}}}{a^{2}}\left[
3-\left( 3+\frac{k^{2}}{\mathcal{H}^{2}}\right) \frac{\left( 5-3\omega _{%
\mathcal{R}}\right) }{\left( 3+3\omega _{\mathcal{R}}\right) }\right] \Phi 
\text{,}  \label{Eq PHi eta b k}
\end{equation}%
where $\mathcal{H}$ can now be expressed as 
\begin{equation}
\mathcal{H}=\frac{H_{0}}{a}\left( \frac{a^{2}+b^{2}}{1+b^{2}}\right) ^{\frac{%
1}{4}}
\end{equation}%
and the initial conditions given by $\Phi \left( a\rightarrow 0\right) =\Phi_{\text{p}}$ and $d\left[ \Phi \left( a\rightarrow 0\right) \right] /da=0$ were assumed. In Fig.~\ref{figPhibk}, we show the complete evolution of potential $\Phi \left( a\right) $ through numerical solutions of Eq.~(\ref{Eq PHi eta b k}). This plot corroborates what we discussed previously, since, for $a\ll b$ (UR behavior), we have a constant potential and one which decays proportional to $\left( k\eta \right) ^{-2}$, for the super-horizon ($k=10^{-3}$ Mpc$^{-1}$) and sub-horizon ($k=1$ Mpc$^{-1}$) modes, respectively. It is also evident from Fig.~\ref{figPhibk} that at $a\gg b$ (NR behavior), the potential is constant in $k$. However, when it crossing from region $a<b$ to $a>b$, the super-horizons modes ($k=10^{-3}$ Mpc$^{-1}$) decrease by $1/10$, so that we have $\Phi \left( a\ll b\right) =(9/10)\Phi \left( a\gg b\right) $. In a realistic universe model, this decrease occurs in the equivalence era \cite{Dodelson,Piattella:2018}. For the hypothetical model considered here, this decrease occurs around $a=b$ (see dotted curves in Fig.~\ref{figPhibk}), where RRG passes from an UR to a NR regime.

It is important to emphasize that, although this model well reproduces the predicted evolution of potential in realistic universes\footnote{In the absence of neutrinos.}, there are no physical reasons for neglecting the quadrupole term (related to anisotropic stress) in the sub-horizon regime since RRG is a non-interacting relativistic gas. Thus, only the results obtained for the super-horizon regime are physically consistent since in this regime we can neglect the quadrupole term. In realistic universe models, the quadrupole term of the photons component can by neglected while they are tightly coupled with the baryons. %In our hypothetical model, the RRG is the only component and does not have self-interaction, so there are no physical reasons which can be used to justify the absence of the quadrupole term when it is relativistic and it is in the sub-horizon regime.

\section{Final comments}\label{final_comments}

In this work we established a theoretical framework for cosmological linear perturbations of RRG described by a distribution function which encodes the main hypothesis of RRG, i.e., that all particles have the same momentum magnitude. We derived the Einstein-Boltzmann system of equations for the model, explicitly verifying that the non-relativistic and ultra-relativistic limits are recovered.

We also studied RRG in the warm approximation, showing that particles with 
$m\sim$ keV can be described as a perfect fluid on scales $k<10$ h/Mpc from now until, at least, $a_{\rm eq}$. For this mass scale, the warm approximation breaks up around $a\sim 10^{-7}$ and the full system of Einstein-Boltzmann equations have to be used. Our results corroborate previous ones \cite{RRG2005BERREDO-PEIXOTO,RRG2018Hipolito,RRG2018MNRAS} which state that RRG is a good approximation to Maxwell-Boltzmann statistics and, therefore, a sufficiently reliable approach to understand and study the behavior of WDM or neutrinos in a model-independent way, without the necessity of solving the full Einstein-Boltzmann system of equations. 

Initial conditions for relativistic RRG in the early universe were also determined and we showed that the evolution of the gravitational potential through radiation and matter dominated eras can be described by a model with RRG only, recovering the general behavior of realistic models.   

It is also interesting to note that the formal description of RRG in terms of a distribution function allows one to include other effects in the Boltzmann equation, such as collisional terms that can model self-interacting dark matter. These models became popular as a possible solution to the Core/Cusp and the Missing Satellites problems \cite{Spergel2000}.

The simple equation of state of RRG motivates the analytical study of relativistic components in cosmological perturbative regime. In particular, it is interesting to look for analytical solutions of density perturbations in a model with WDM, radiation and $\Lambda $, where WDM is modeled by RRG. This kind of study involving generalizations of M\'{e}sz\'{a}ros equation \cite{meszaros1974behaviour} is being carried out by the authors.

%\appendix
%\section{Some title}
%Please always give a title also for appendices.
\acknowledgments

G. Pordeus da Silva thanks CNPq-Brazil (141165/2017-0), government of the state of
Para\'{\i}ba-Brazil and CAPES-Brazil for the financial support.

\bibliography{MyReference}

\end{document}